\documentclass[preprint,showpacs,preprintnumbers,amsmath,amssymb]{revtex4}

\usepackage{graphicx}
\usepackage{dcolumn}
\usepackage{bm}

\begin{document}

\preprint{APS/123-QED}

\title{Shock wave propagation in vibrofluidized granular materials}

\author{Kai Huang}%
 \email{huangkai1996@nju.org.cn}
\author{Guoqing Miao}%
 \email{miaogq@nju.edu.cn}
\author{Yi Yun}%
\author{Hua Zhang}%
\author{Rongjue Wei}
 \affiliation{State Key Laboratory of
Modern Acoustics and Institute of Acoustics, Nanjing University,
Nanjing 210093, P. R. China}

\date{\today}

\begin{abstract}
Shock wave formation and propagation in two-dimensional granular
materials under vertical vibration are studied by digital high
speed photography. The steepen density and temperature wave fronts
form near the plate as granular layer collides with vibrating
plate and propagate upward through the layer. The temperature
front is always in the transition region between the upward and
downward granular flows. The effects of driving parameters and
particle number on the shock are also explored.
\end{abstract}

 \maketitle

Granular materials are ubiquitous in nature and play an important
role in many of our industries and daily lives \cite{NagelReview}.
Due to their noncohesive, strongly dissipative properties,
granular materials behave differently from usual solids, liquids
and gases. Under vertical vibrations, as the input energy
increases, the state of granular materials changes from
solid-like, to liquid-like and gas-like state. The three states
and phase transitions between them have been studied
experimentally and theoretically \cite{Goldshtein95}. Fluidized
granular materials show many interesting phenomena, such as
surface pattern \cite{Melo}, oscillon\cite{Umbanhowar}, convection
\cite{Nagel_earlier}, size and density segregation
\cite{Knight93}, cluster \cite{cluster}, heap formation
\cite{Faraday1883Evesque89Nagel93} and transport \cite{hkai}.
Recently sound wave propagation attracts much interest because it
coexist with most of the above phenomena and is not well
understood. It is found that sound in close packed granular
materials mainly propagates through force-chains. The speed of
sound in sand is approximately the same as that in air and very
sensitive to the arrangement of grains \cite{Cheng}. Computer
simulations indicate that there exist density and pressure waves
in granular materials under vertical vibrations \cite{Aoki95}.
Recently time-independent shocks in granular flow past an obstacle
are observed by experiments and compared with simulation results
\cite{StaticShock}. Moreover the formation and propagation of
shocks in vibrofluidized granular materials have been studied by
molecular dynamic simulations and numerical analysis of continuum
equations \cite{shock2002}. However experimental study of wave
propagations in vibrofluidized granular materials is scarce. In
this paper we use high speed photography to explore shock
formation and propagation in vertical vibrated bidimensional
granular materials.

The experiment is conducted with a rectangular container mounted
on the vibrating exciter (Br\"{u}el \& Kj$\ae$r 4805), which is
controlled by a function generator (type HP 3314A). We use steel
spheres with diameter $d=4$ $mm$ and density $\rho=7900$ $kg/m^3$.
The container is made up of two parallel, $l=90mm$(length) and
$h=280mm$(height), glass plates separated by $w=4.1$ $mm$
vertically adhered in a Plexiglas bracket. The total particle
number $N$, the driving frequency $f$ and the nondimensional
acceleration $\Gamma = 4 \pi^{2} f^{2}A/g$ ($A$ is the driving
amplitude and $g$ the gravitational acceleration) are used as
control parameters. A high speed camera (Redlake MASD MotionScope
PCI 2000sc) is used to record the movements of the spheres. A
frequency multiplication and phase lock circuit is used to
generate external trigger signals for the camera. The acquisition
rate is $N_p$$\times$$f$ with multiple number $N_p=25$. Every
recorded image is $22d$(length)$\times$$30d$(height). Image
processing technique \cite{wildman2000warr95powder942000} is used
to track the locations of all particles. We use vibrating
container as reference frame in image processing. Every image of
grains is divided vertically into a number of strips, each with
width of $1d$ and indicated by $1, 2, ...$ from bottom to top.
Assembly average is performed over $600$ cycles for granular
properties at $N_p$ phase points in each cycle. Granular density
$D(jd,k\Delta t)$ is the number of particles in $jth$ strip at
time $k\Delta t$, in which $\Delta t=1/(N_pf)$. For brevity, we
omit the units of space and time and use $D(j,k)$ instead of
$D(jd,k\Delta t)$ hereafter. Granular temperature is defined by
$T(j,k)=\sum^{N_s}_{n=1}\frac{1}{2}\vert {\bf v}_n-{\bf v}_b(j,k)
\vert^2/N_s$, in which $N_s$ is the number of particles in $jth$
strip at time $k$, ${\bf v}_n$ is the velocity vector of the $nth$
particle, and the background velocity ${\bf
v}_b(j,k)=\sum^{N_s}_{n=1}{\bf v}_n/N_s$ is the mean particle
velocity in $jth$ strip at the time $k$ averaged over all cycles
recorded.

\begin{figure}
    \centering
    \begin{minipage}[b]{0.25\textwidth}
    \includegraphics[width=0.9\textwidth]{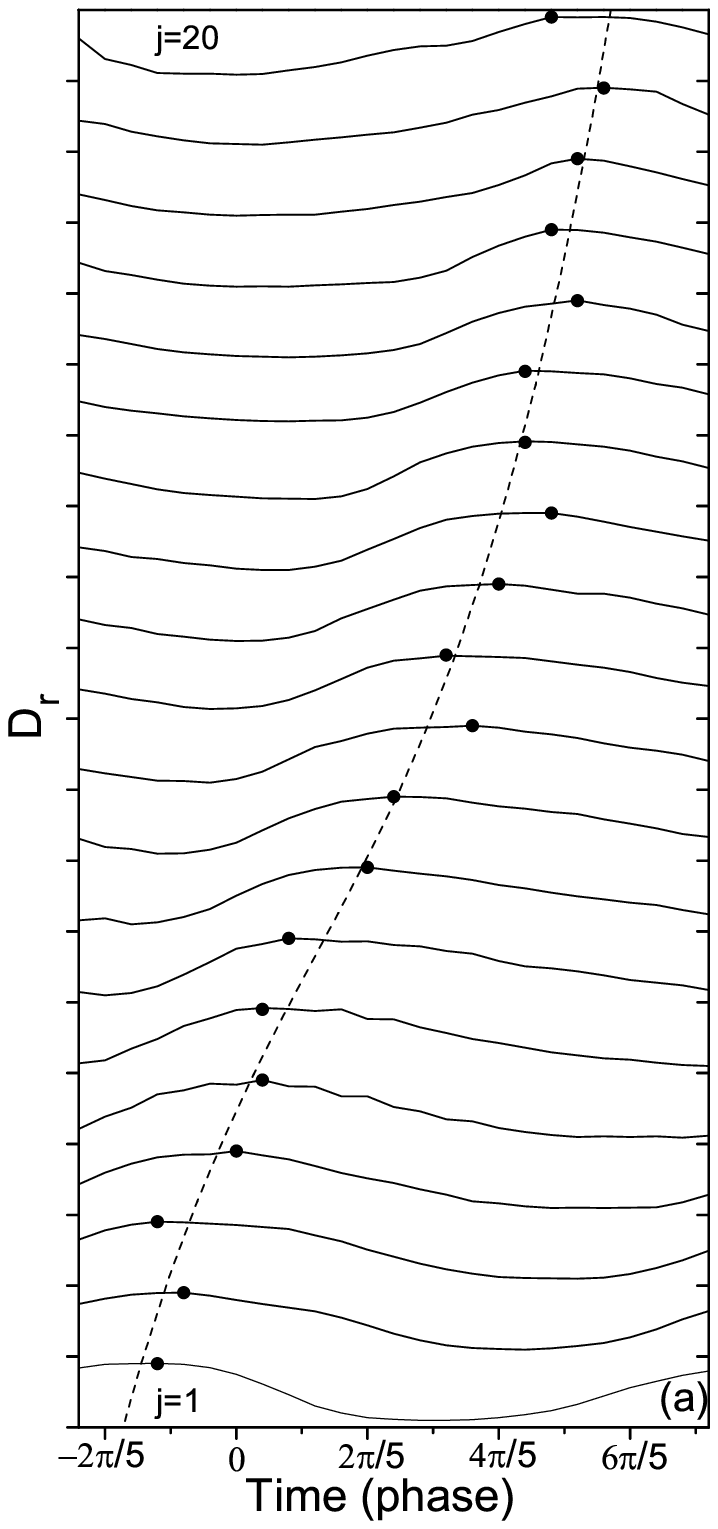}
    \end{minipage}%
    \begin{minipage}[b]{0.25\textwidth}
    \includegraphics[width=0.9\textwidth]{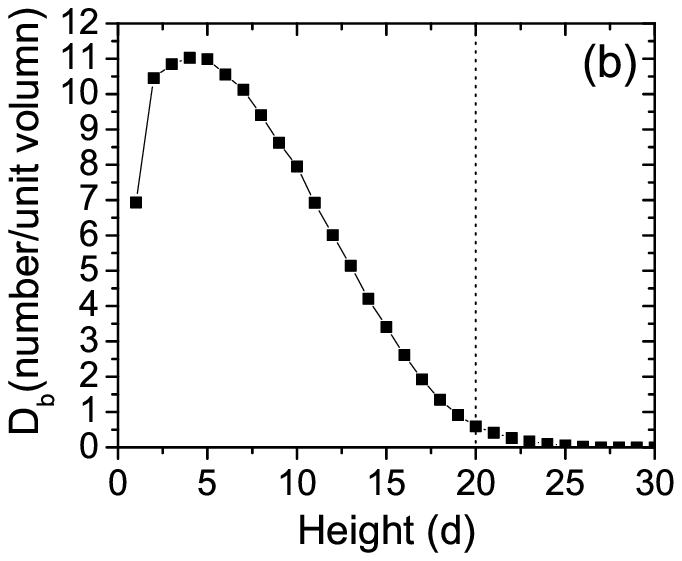}
    \end{minipage}
\caption{\label{fig:density}(a) Time-space profiles of density
wave $D_r(j,k)$ from $1st$ to $20th$ strips, solid circles peak
points of density at each strip, dashed line polynomial fitting of
peak values with time; time $=0$ corresponds to the time when the
plate reaches its maximum height; (b) background density $D_{b}$
versus height, which is below $1$ at height above strip $20$
(dotted line). Parameters are $f=15Hz$, $\Gamma=5$ and $N=150$.
The height of peak point at each strip is set at 90 percent of
total height of that strip.}
\end{figure}

As $\Gamma$ increases to and beyond $1$, granular layer fluidizes
from upper to lower parts and the density fluctuations or density
wave defined by $D_r(j,k)=D(j,k)-D_b(j)$ (the background number
density $D_b(j)=\sum^{N_p}_{k=1}D(j,k)/N_p$) appears in the layer
(as shown in Fig.~\ref{fig:density}(a)). The bottom of the layer
reaches its maximum density at time $=-\pi/5$.
Fig.~\ref{fig:density}(a) shows the time-space profile of density
wave. The fitting curve of density peaks of all the strips
indicates that the density wave propagates upward with a
nonuniform velocity, faster below than above $4th$ strip and is in
agreement with the results obtained from molecular dynamic
simulations \cite{Aoki95}. The wave distorts as it moves upward.
It is well known that in ordinary gas the distorted wave will
evolve into shock wave as it propagates. The calculation of Mach
number $Ma$ indicates that there exists shock wave in our
experiment. As a supersonic granular flow encounters an
impenetrable plate, velocities of particles (relative to the
plate) near the plate abruptly decrease to about zero, but those
of undisturbed particles are still unchanged. This results in a
normal shock formation near the plate. Then this shock wave leaves
plate and propagates upward. We only plot density profiles below
$20th$ strip because above that height the number density $D_b$ is
too small for accurate statistics (Fig.~\ref{fig:density}(b)).

\begin{figure}
    \begin{minipage}[b]{0.45\textwidth}
    \includegraphics[width=.85\textwidth]{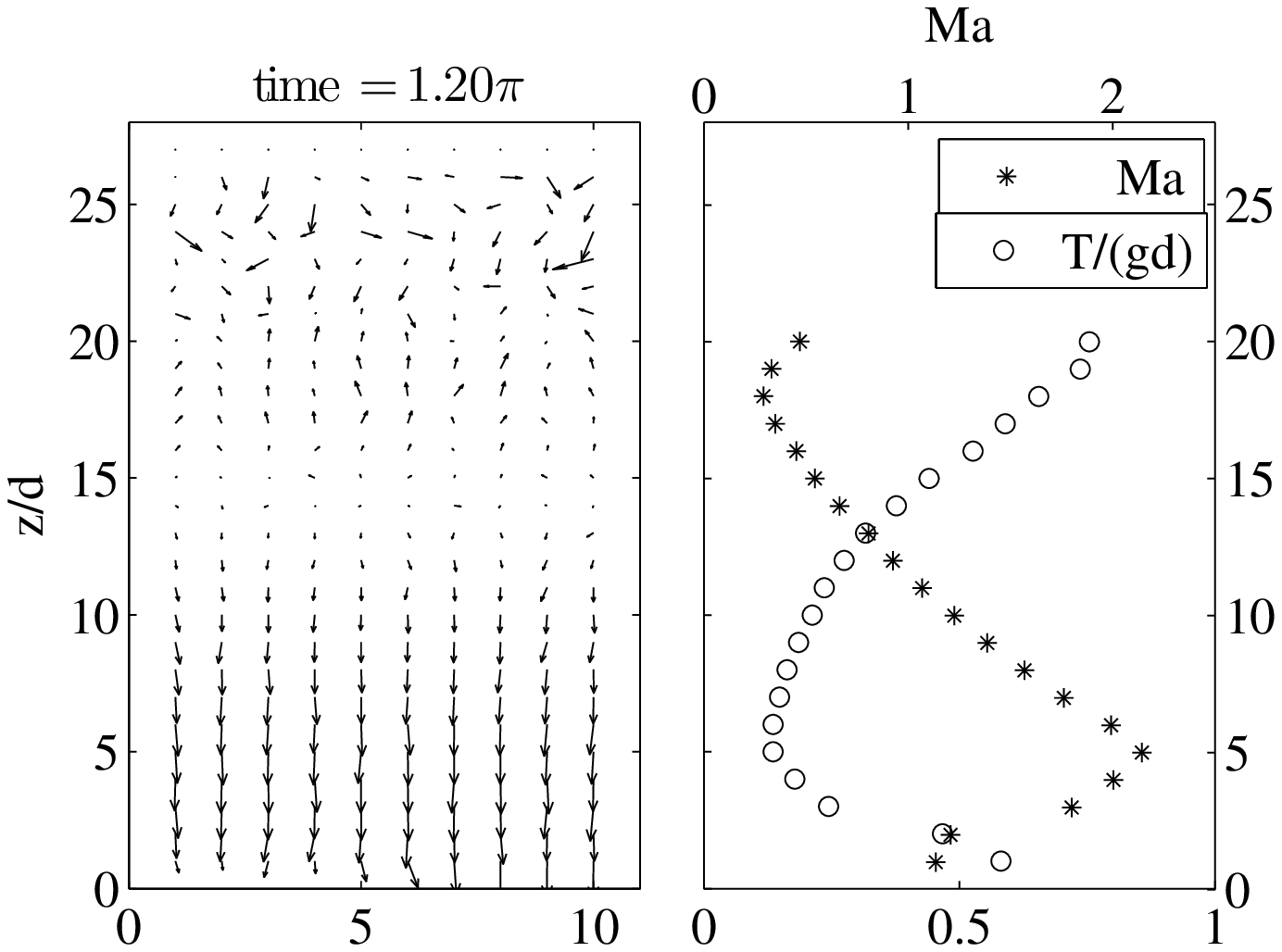}
    \includegraphics[width=.85\textwidth]{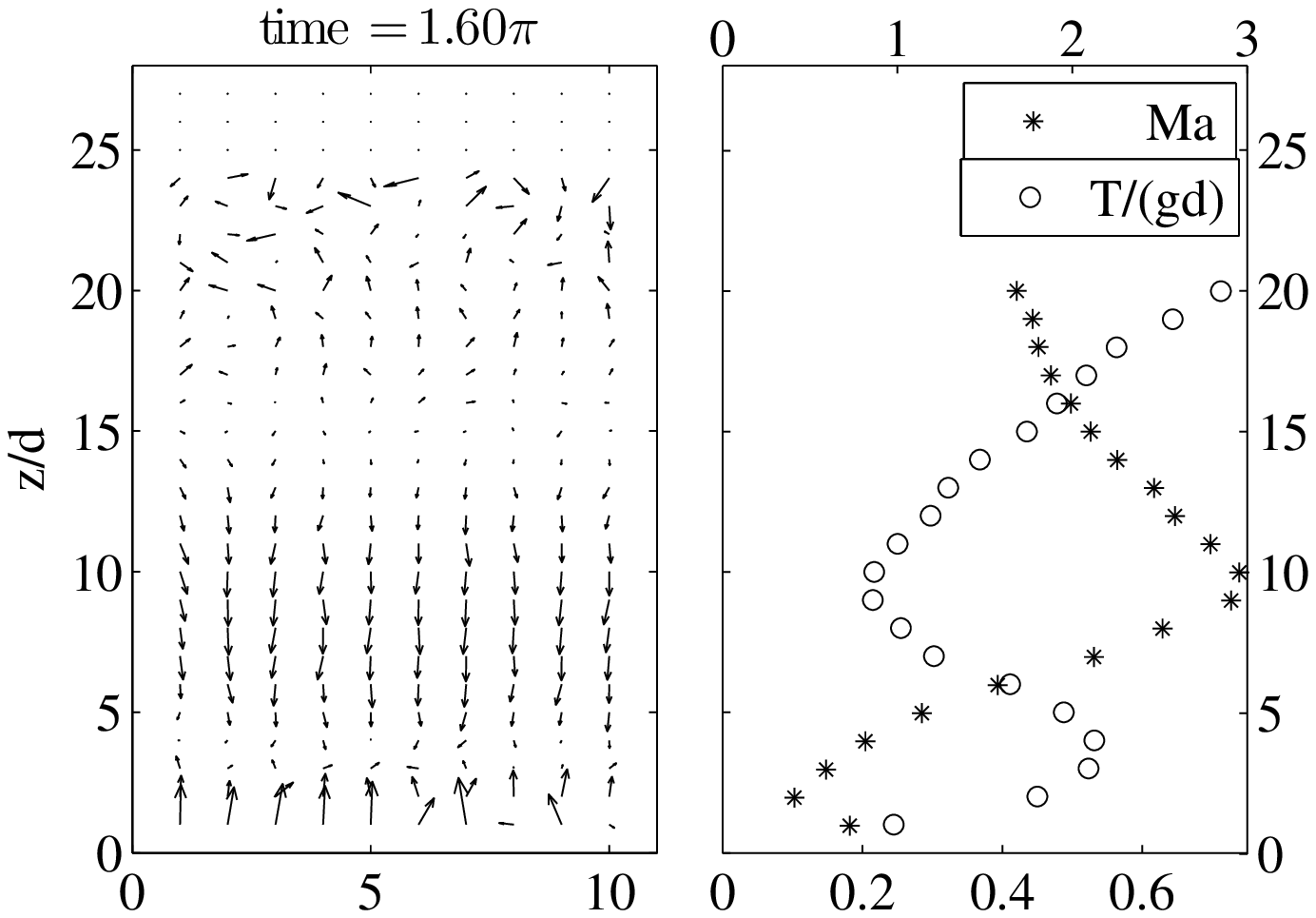}
    \includegraphics[width=.85\textwidth]{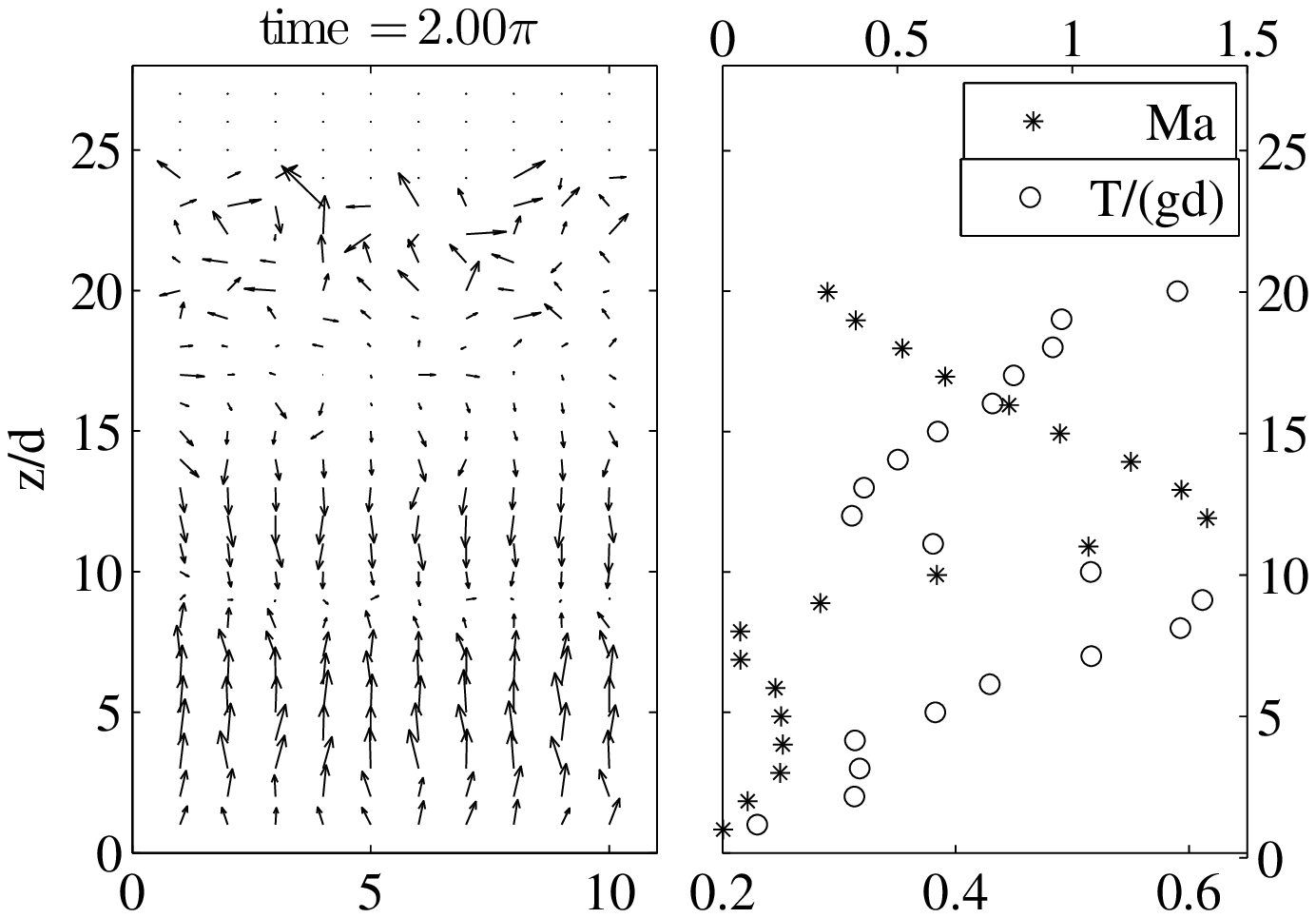}
    \includegraphics[width=.85\textwidth]{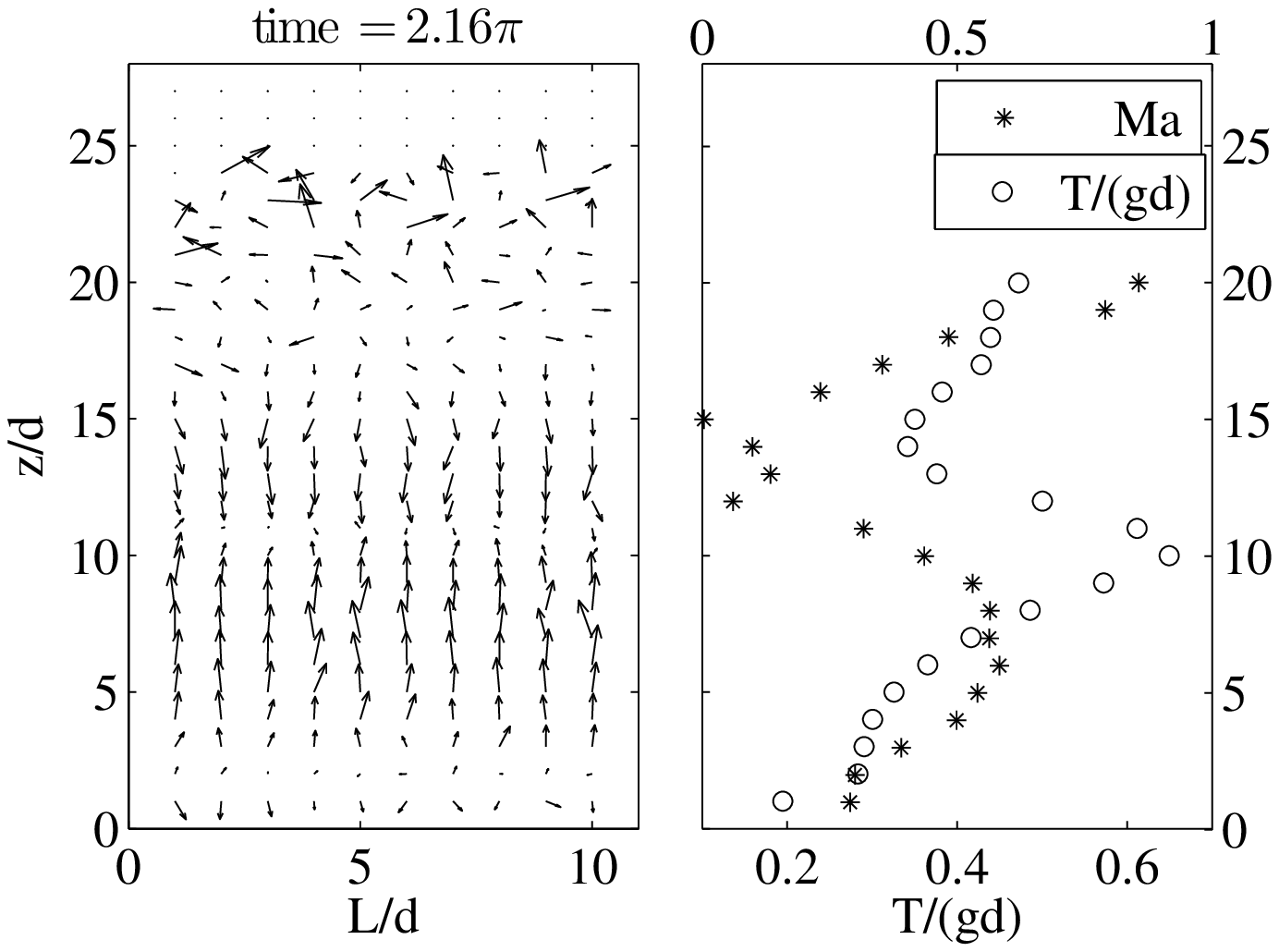}
    \end{minipage}
\caption{\label{fig:T} Velocity field (left column), Mach number
$Ma$(asterisk) and scaled temperature(open circle)(right column)
as a function of height(right) at four different times. At time
$=\pi$ the plate is at its lowest position. Parameters here are
the same as those in Fig.\ref{fig:density}.}
\end{figure}

\begin{figure}
    \begin{minipage}[b]{0.45\textwidth}
    \includegraphics[width=.85\textwidth]{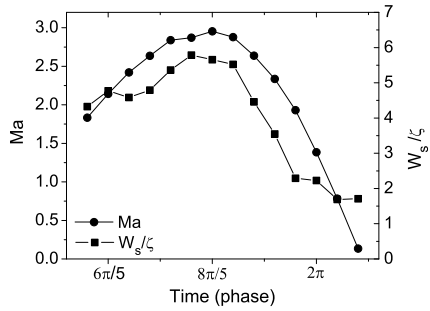}
\end{minipage}
\caption{\label{fig:Ma} The peak value of Mach number and shock
width $W_s/\zeta$ relative to the mean free path $\zeta$ as a
function of time. The shock begins to form at time$=1.20\pi$ and
disappear at $2.00\pi$. The shock width decreases to about $1.5$
at time$=2.00\pi$, indicating that the shock is fully developed.
Parameters here are the same as those in Fig.~\ref{fig:density}.}
\end{figure}

In the experiment the shock is identified in a region where the
Mach number increases from its minimum ($<1$) in the disturbed
region to the maximum ($>1$) in the undisturbed region. The Mach
number is defined by

\begin{equation}
Ma=\vert\frac{v_{bn}-v_p}{c}\vert,
\end{equation}

\noindent where ${v_{bn}}$ is the vertical component of ${\bf
v_b}$, $v_p$ is the velocity of the plate and $c$ is the speed of
sound calculated with \cite{shock2002,Savage88},

\begin{equation}
c=\sqrt{T\chi(1+\chi+\frac{\nu}{\chi}\frac{\partial{\chi}}{\partial{\nu}})},
\end{equation}

\noindent in which
$\chi=1+2(1+e)\nu[1-({\nu}/{\nu_{max}})^{{4\nu_{max}}/{3}}]^{-1}$,
$e$ is the restitution coefficient, $\nu(j,k)=D(j,k)\pi d^2/(6lw)$
is the volume fraction and $\nu_{max}=N\pi d^3/(12h_{cm}lw)=0.57$
is the maximum volume fraction \cite{wildman2000warr95powder942000}. The height $h_{cm}$ of the center
of mass at rest is calculated by

\begin{equation}
h_{cm}=\frac{n_bd}{2N}[(1-\sqrt{3}/2)n_h+(\sqrt{3}/2)n_h^2]+\frac{n_0d}{2N}(1+\sqrt{3}n_h),
\end{equation}

\noindent in which $n_b=l/d-0.5$ is the average number of
particles per strip, $n_h=int(N/n_b)$ is the number of full strip
and $n_0=N-n_hn_b$ is the particle number at the highest strip
\cite{wildman2000warr95powder942000}. We investigate the
propagation of the shock by examining the distribution of granular
density, temperature and Mach number at four time in one cycle
(shown in Fig.~\ref{fig:T}). The maximum Mach number in the shock
and the shock width relative to the mean free path of particles
$\zeta=lw/(\sqrt{8\pi}dD)$ \cite{shock2002} as a function of time
are drawn in Fig.~\ref{fig:Ma} to show the shock dynamics.

At time $=1.20\pi$, the plate moves upward with $v_p=0.306m/s$ and
begins to collide with the granular layer. Particles at lower
strips begin to be compressed by collision. In the mean time most
of the particles in the undisturbed region are still falling
towards the plate with supersonic speed. This results in the
formation of a shock wave. In the shock region the Mach number
increases with height and reaches a maximum value $2.73$ at the
$5th$ strip. The particle velocities are randomized by collision
and the granular temperature decreases with height in this region.

At time $=1.60\pi$, the plate moves upward with $v_p=0.495m/s$.
The granular layer continues to be compressed on the plate, more
and more particles come into the compressed region, resulting in
the propagation of the shock up through the layer. In the shock
region, the Mach number increases with height and reaches its
maximum ($Ma=2.95$) at $10th$ strip (Fig.3). A temperature peak
appears at $3rd$ strip which is the bottom of shock region and is
the transition region between upward and downward granular flows.

At time $=2.00\pi$, the plate reaches its maximum height. The
granular layer leaves the plate and flies freely. The shock
propagates upward with the maximum Mach number decreasing to
$1.39$ at the $12th$ strip. As seen in Fig.~\ref{fig:Ma}, the
shock width now decreases to its minimum (approximate $1.7$ times
the mean free path), indicating that the shock is fully developed.
The steepen temperature wave front propagates upward to $10th$
strip.

After time $=2.00\pi$, the plate begins to move downward. The
maximum Mach number in the shock region decreases. At time
$=2.16\pi$, the shock disappears because the maximum Mach number
at strip $13$ damps to be less than unity. The temperature wave
front continues to propagate upward until disappears in the dilute
region of the layer. Then the granular layer is ready for the next
collision with the plate.

\begin{figure}
    \begin{minipage}[b]{0.45\textwidth}
    \includegraphics[width=.85\textwidth]{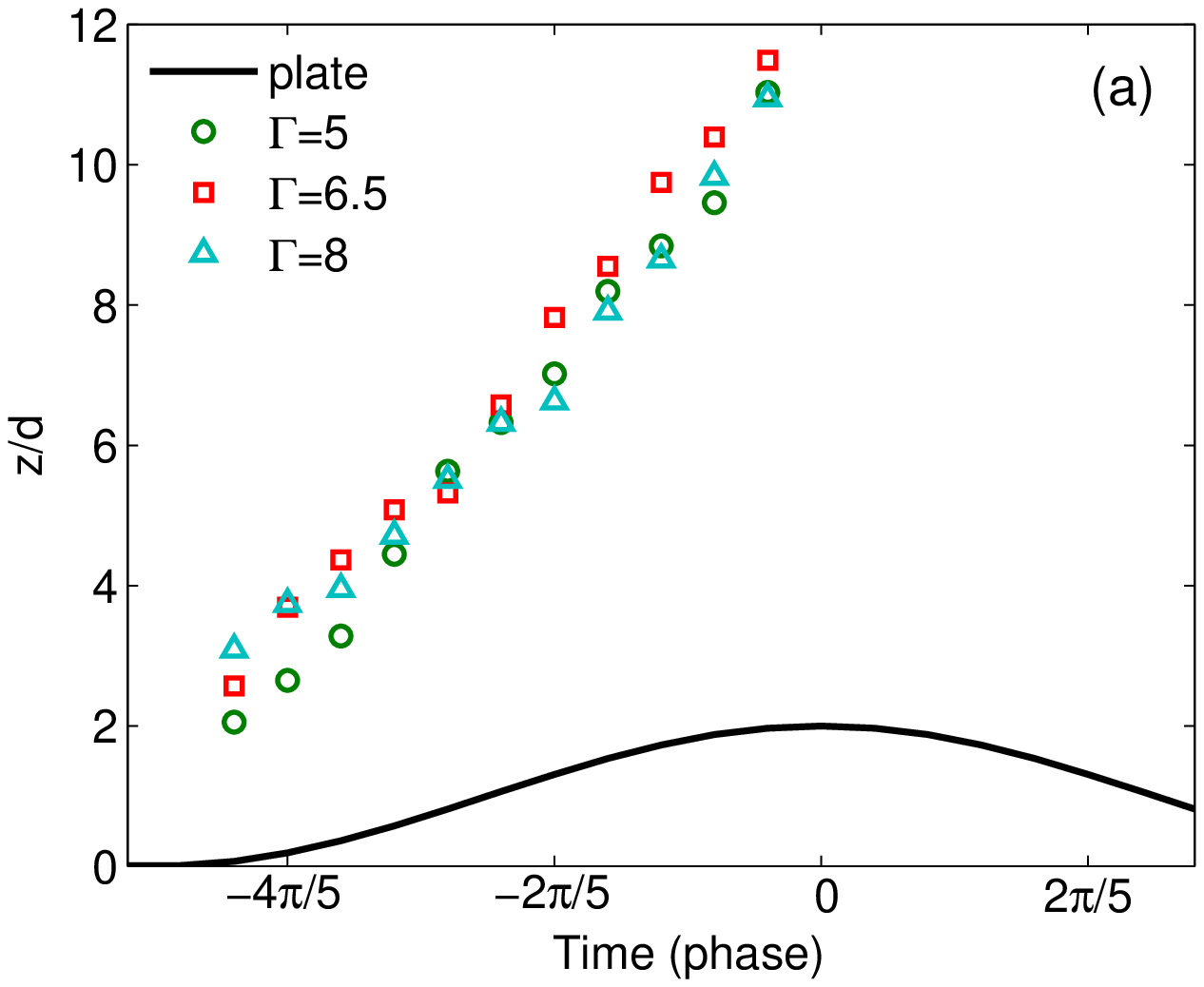}
    \includegraphics[width=.85\textwidth]{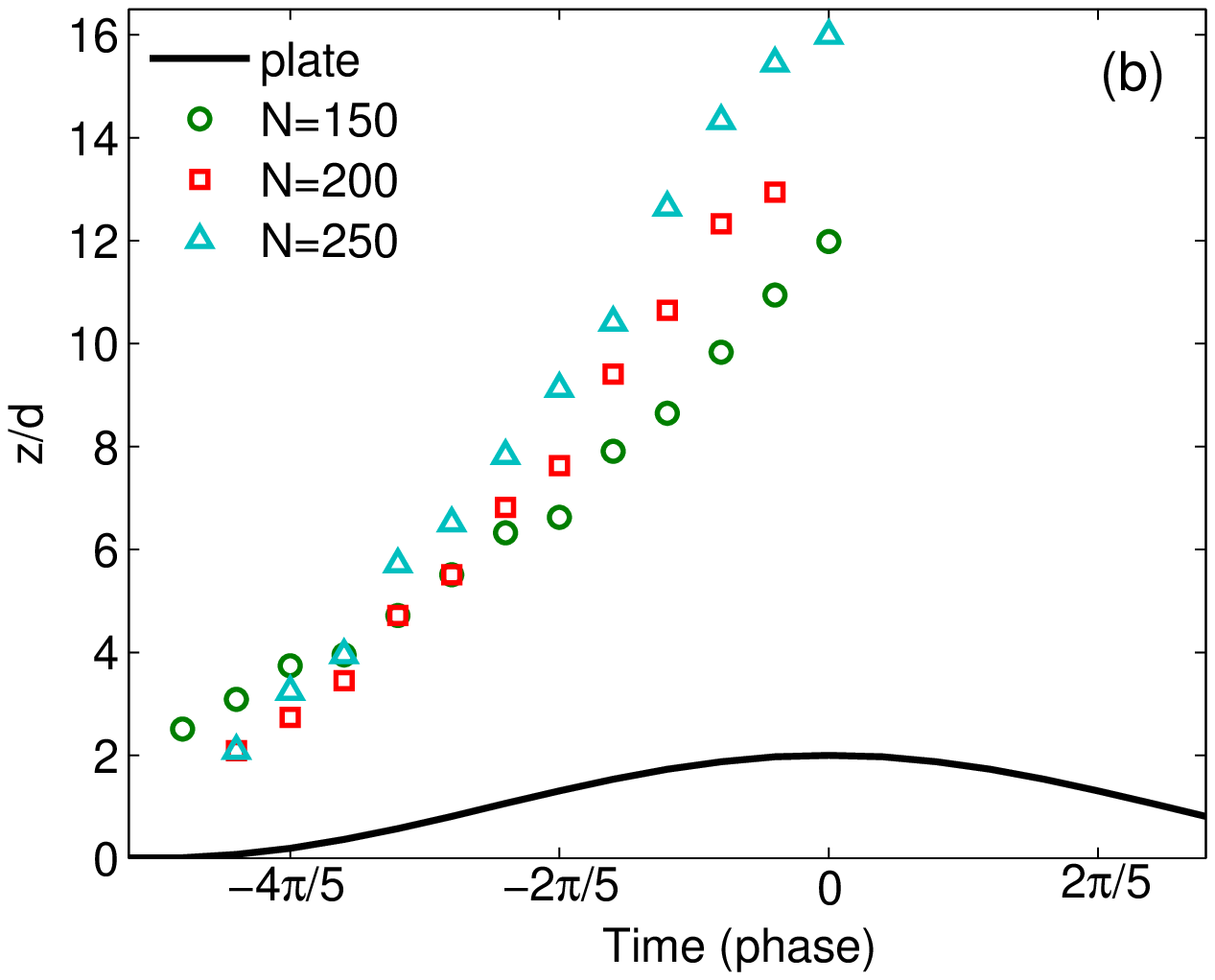}
    \end{minipage}
\caption{\label{fig:track} Shock location as a function of time.
with different $\Gamma$ (a) and $N$ (b). Other parameters are
fixed at $f=20Hz$, $N=150$ in (a) and $f=20Hz$, $\Gamma=8$ in (b).
Solid line corresponds to the vibrating plate.}
\end{figure}

To investigate how changing $\Gamma$ and $N$ affect the shock, we
perform the experiment with $\Gamma$ from $5$ to $8$ and $N$ from
$60$ to $250$. As $\Gamma$ increases, the velocity of granular
layer relative to the plate increases, resulting in the increase
of maximum Mach number and the speed of shock. On the other hand,
as $\Gamma$ increases, the average number density decreases
(particles expand to more space of the container), and this
results in a decrease of the propagating velocity of the wave
\cite{Harada}. As a result the total effects of these two opposite
factors make a little change of the velocity of shock with the
change of $\Gamma$ (Fig.~\ref{fig:track}(a)). As total number of
particles increases from $150$ to $250$ (the layer depth from $7d$
to $11d$) with driving parameters fixed, the average number
density increases. As indicated in Fig.~\ref{fig:track}(b), denser
layer collides with the plate later than more dilute layer. Thus
increasing layer depth causes the shock to form later in the cycle
and the shock propagates faster through the higher density layer.

In conclusion, we prove the existence of shock wave in
vibrofuidized granular materials experimentally. The shock forms
as particles collide with vibrating plate and propagates upward
with a steepen temperature front in the transition region between
upward and downward granular flows. The velocity of shock depends
on the velocity of the plate when it collides with granular layer
and the number density of granular layer. This is helpful for the
understanding of phenomena in vibrofluidized granular materials
such as surface instability, convection and energy transfer, etc.

This work was supported by the Special Funds for Major State Basic
Research Projects, National Natural Science Foundation of China
through Grant No. 10474045 and No. 10074032, and by the Research
Fund for the Doctoral Program of Higher Education of China under
Grant No. 20040284034.

\end{document}